\documentclass[aps,prl,reprint,superscriptaddress,nofootinbib,longbibliography]{revtex4-2}
\usepackage{amsmath,amssymb,bm}
\usepackage{microtype}
\usepackage[colorlinks=True]{hyperref}
\newcommand{\A}{\mathcal A}

\newcommand{\Hmax}{H_{\max}}
\newcommand{\Hmin}{H_{\min}}
\newcommand{\Hgen}{H_{\max,\mathrm{gen}}}
\newcommand{\eps}{\varepsilon}
\newcommand{\Int}{\operatorname{Int}}

\begin{document}
\title{Entropy Obstruction to Closed Semiclassical Bounces}
\author{Naman Kumar}
\affiliation{Department of Physics, Indian Institute of Technology Gandhinagar, Gujarat 382355, India}
\email{naman.kumar@iitgn.ac.in}
\date{\today}
\begin{abstract}
We prove a finite-$G\hslash$ singularity theorem for semiclassical spacetimes with compact Cauchy slices. Let a compact Cauchy slice be divided by a compact surface into regions $B$ and $C$. Suppose that $B$ is conditionally hyperentropic, $\Hgen^\eps(BC|C)>0$, that the future-inward null boundary from the dividing surface toward $B$ is a discrete max lightsheet, and that $C$ is robustly quantum trapped. Assuming discrete max-focusing and a regular semiclassical endpoint for a closing lightsheet, the future Cauchy development of $B$ contains an incomplete null generator. This parallels entropy singularity theorems in which hyperentropy is combined with the existence of an inward lightsheet, while robust quantum trapping supplies the additional finite-$G\hslash$ local obstruction needed here. In a closed Friedmann universe, the theorem excludes a controlled semiclassical bounce when a contracting hemisphere lies on such a lightsheet and contains more independent information than its boundary can support.
\end{abstract}

\maketitle

\textit{Introduction.--}
Singularity theorems do not identify a point at which curvature must diverge or spacetime must literally break down. Their conclusion is more modest: at least one causal geodesic ends after a finite value of its affine parameter, so the spacetime is geodesically incomplete. This incompleteness is then conventionally promoted to the rather dramatic name ``singularity''. The terminology is therefore somewhat grander than the mathematical conclusion itself. Still, geodesic incompleteness does signal that the spacetime description cannot be continued indefinitely within the assumed framework, so some genuine pathology is present even if its precise nature is left unspecified. Penrose obtained this conclusion from a trapped surface, the null energy condition, global hyperbolicity, and the existence of a noncompact Cauchy slice \cite{Penrose1965}. The Hawking--Penrose theorems and later kinematical results broadened the range of situations in which geodesic incompleteness can be established, including cosmological settings \cite{HawkingPenrose1970,BordeGuthVilenkin2003}. Together, these results explain why avoiding geodesic incompleteness in expanding or collapsing cosmologies is highly nontrivial.

A trapped surface is a closed surface for which both future-directed families of orthogonal null rays initially converge. Raychaudhuri's equation \cite{Raychaudhuri:1953yv}, together with the null energy condition (NEC), then forces these rays to focus within a finite affine distance. This provides the local part of Penrose's argument. The global step compares the resulting compact null boundary with the assumed noncompact Cauchy slice and obtains a contradiction. This distinction is essential for closed universes: there, the null boundary may simply close on itself without conflicting with the topology of the compact Cauchy slice.

Two assumptions become problematic when one asks whether quantum effects can replace the big bang or big crunch by a nonsingular bounce. First, the null energy condition can be violated by physically admissible quantum states. This is not merely a technical complication: the negative energy flux associated with Hawking radiation is essential to black-hole evaporation \cite{Hawking1975}. Second, a closed universe has compact Cauchy slices, so the topological contradiction used in Penrose's argument is no longer available. Global de Sitter space makes this clear: its contracting phase contains trapped two-spheres, yet the full spacetime remains null geodesically complete.

This gap is directly relevant to closed bounce models. During contraction, a sufficiently large sphere can become trapped even if the scale factor later reaches a smooth minimum. The classical compact-Cauchy argument therefore does not by itself exclude a nonsingular bounce. The situation changes when one side of the trapped sphere contains more independent quantum information than the sphere can support as gravitational entropy. A complete contraction followed by re-expansion would then require this information to pass through a closing null boundary whose entropy capacity is too small. This suggests that, in a compact universe, excess quantum information can replace the noncompactness assumption that drives the global part of Penrose's argument. The theorem developed below makes this statement precise by combining a conditional information bound with a suitable inward lightsheet and robust quantum trapping. Thus even a low-curvature bounce can fail if a contracting closed geometry is accompanied by an entropy-rich, weakly correlated matter state. Vacuum-like states, including global de Sitter, need not satisfy the required information condition.

This information-based global obstruction still requires a quantum replacement for the local focusing step in Penrose's argument. Black-hole thermodynamics provides a natural candidate. Bekenstein's generalized entropy combines the horizon area with the entropy of quantum fields outside the horizon \cite{Bekenstein1973,Bekenstein1974}. The generalized second law (GSL) states that this total entropy does not decrease along a causal horizon, and it has been proved under broad field-theory assumptions for arbitrary horizon cuts \cite{WallGSL2012}. Wall then used the GSL to formulate a quantum singularity theorem in which the classical null expansion is replaced by the variation of generalized entropy \cite{Wall2013}.

The covariant entropy bound supplies the complementary global ingredient. Instead of constraining entropy by a spatial volume, it bounds the entropy associated with null hypersurfaces of nonpositive expansion \cite{Bousso1999,Bousso2002,FlanaganMarolfWald2000}. Quantum versions replace the classical area decrease by a decrease of generalized entropy \cite{StromingerThompson2004,BoussoCasiniFisherMaldacena2014}. This line of work led to the quantum focusing conjecture \cite{BoussoFisherLeichenauerWall2016} and to singularity theorems based on hyperentropic or hyperentangled regions \cite{BoussoShahbazi2022,BoussoShahbazi2023}.

There is, however, an important finite-$G\hslash$ issue. Wall's quantum singularity theorem uses a quantum analogue of the classical touching argument, whose control is tied to the regime in which the gravitational area term dominates, effectively $G\hslash\to0$ \cite{Wall2013}. In a genuinely semiclassical regime at finite $G\hslash$, the area and matter-entropy variations can be comparable, and a pointwise touching comparison need not be valid. The robust singularity theorem was designed precisely to avoid this limitation by replacing the pointwise comparison with a discrete, information-theoretic formulation that remains meaningful at finite $G\hslash$ \cite{Bousso2025}. The remaining gap is then clear: entropy singularity theorems can accommodate compact Cauchy slices, while the robust finite-$G\hslash$ theorem supplies the needed local control but retains a spatial-openness assumption. 

At finite $G\hslash$, this limitation becomes important because the area and matter-entropy contributions to the quantum expansion can be comparable. Two distinct surfaces therefore cannot, in general, be compared simply because they touch at a point. One-shot entropy and discrete max-focusing avoid this difficulty by comparing nested spacetime regions rather than pointwise quantum expansions \cite{AkersEtAl2024,BoussoTabor2025}. The same framework also provides one-shot versions of the GSL~\cite{AkersEtAl2024} and connects naturally with the quantum Bousso-bound and quantum-focusing literature~\cite{Bousso1999,StromingerThompson2004,BoussoCasiniFisherMaldacena2014,BoussoFisherLeichenauerWall2016}.

In this Letter we use these ideas to extend the robust finite-$G\hslash$ singularity framework to compact Cauchy slices. Following entropy-based singularity theorems, we assume both excess independent information in $B$ and the existence of an inward entropy-bounded null sheet, here formulated as a discrete max lightsheet. Robust quantum trapping plays a separate local role: it excludes a complete generator that could remain indefinitely on the relevant Cauchy horizon without requiring a pointwise comparison of quantum expansions. If all generators were nevertheless complete, compactness would force the horizon to close. The max-Bousso inequality then applies to every finite cut of the assumed lightsheet, and regular endpoint continuity carries this inequality to the limiting empty-edge wedge, contradicting $\Hgen^\eps(BC|C)>0$. The lightsheet assumption and robust trapping therefore perform distinct roles in the proof. 

The Letter also includes End Matter that provides further technical details on the localized wedge construction, the compactness argument, and the regulated endpoint limit. It also gives proofs of several auxiliary results used in the main text.
\newpage
\textit{Initial data and information condition.--}
Let $(M,g,\rho)$ be an inextendible, globally hyperbolic semiclassical spacetime admitting Cauchy slices on which the gravitational effective theory is valid. Let a compact Cauchy slice $\Sigma$ be partitioned as
\begin{equation}
 \Sigma=B\cup\sigma\cup C,\qquad \sigma=\partial_\Sigma B=\partial_\Sigma C,
 \label{partition}
\end{equation}
where $B$ and $C$ have nonempty interiors and $\sigma$ is a compact $C^1$ surface. Here $B$ and $C$ are understood together with their common edge $\sigma$ whenever they are used as initial regions in domains of dependence or Cauchy horizons. Thus, for example, $D^+(B)$ and $H^+(B)$ refer to the domain of dependence and future Cauchy horizon of the closed initial region $B\cup\sigma$. The future null direction from $\sigma$ into $B$ is equivalently the future outward null direction of $C$.

For nested regions, one-shot quantum information is measured by a smooth conditional max entropy. Unlike the von Neumann entropy, this quantity does not assume that many identical copies of the state are available. It asks how many effective states are needed to describe one region when the state of a second region is known, allowing an error controlled by the smoothing parameter $\eps$. The conditional quantity $\Hmax^\eps(B|C)$ therefore measures the information in $B$ that cannot be inferred from $C$. In the many-copy limit it approaches the corresponding von Neumann conditional entropy, but that limit is not used here \cite{AkersEtAl2024}.

In a gravitating theory, the information carried by the fields must be combined with the entropy of the boundary. The generalized conditional max entropy is \cite{AkersEtAl2024,BoussoTabor2025}
\begin{equation}
 \Hgen^\eps(X|Y)=\frac{\A(X)-\A(Y)}{4G\hslash}+\Hmax^\eps(X|Y)+O(G\hslash).
 \label{generalizedmax}
\end{equation}
Here $\A$ denotes the area of the edge of the corresponding spacetime region. The symbol $O(G\hslash)$ also includes the local higher-curvature counterterms required by the gravitational effective action. The area and matter terms are separately regulator dependent, but their generalized sum is finite\footnote{For example, the leading ultraviolet divergence of the vacuum entanglement entropy across the surface $\sigma$ scales as $A(\sigma)/\epsilon_{\rm UV}^{2}$ in four spacetime dimensions. This divergence is absorbed into the renormalization of $1/G$, so that the corresponding divergence in the matter entropy is cancelled by the area term. Subleading ultraviolet divergences are likewise absorbed by the renormalization of higher-curvature couplings in the gravitational effective action. Thus the separate area, matter-entropy, and counterterm contributions are regulator dependent, while their renormalized generalized-entropy combination is finite.}. Since the complete Cauchy slice has no boundary while $\partial C=\sigma$,
\begin{equation}
 \Hgen^\eps(BC|C)=-\frac{\A(\sigma)}{4G\hslash}+\Hmax^\eps(B|C)+O(G\hslash).
 \label{hypercondition}
\end{equation}
The condition $\Hgen^\eps(BC|C)>0$ says that the information in $B$ that is not predictable from $C$ exceeds the gravitational entropy carried by the common boundary. We call such a partition \emph{conditionally hyperentropic}. This differs from requiring a large thermodynamic entropy in $B$. If $B$ is strongly entangled with $C$, much of its local entropy may already be encoded in $C$ and does not contribute as independent information. The conditional inequality is therefore the appropriate statement for a compact universe, where there is no external system that can be ignored.

An elementary analogy is data compression with side information. A file may look complicated by itself but require few additional bits when a correlated file is already available. Likewise, \(B\) can have a large entropy even when much of its information can already be recovered from \(C\). The theorem counts only the part that is not determined by \(C\). This quantity is also unchanged if the spatial slice is moved within the same domains of dependence, since unitary evolution can rearrange the information but cannot create new independent information behind the null boundary.

The second condition is geometric. To formulate it without using a pointwise quantum expansion, we identify each spatial region with its domain of dependence. Let $a$ denote such a region. We say that $a$ is \emph{past-nonexpanding} at an edge point $p$ if, within some open neighborhood of $p$, every allowed past-directed outward null enlargement $b\supset a$ satisfies
\begin{equation}
\Hgen^\eps(b|a)\leq0.
\label{pne}
\end{equation}
This is the discrete entropy analogue of nonnegative future outward expansion. The region $a$ is \emph{past-noncontracting} if no sufficiently small, proper, future outward null deformation is past-nonexpanding at all of its new edge points. Finally, $a$ is \emph{robustly quantum trapped} if its edge is compact and it is past-noncontracting \cite{Bousso2025}. In the regime where the area term dominates, a surface with strictly negative classical future expansions satisfies this condition after an arbitrarily small generic deformation. The discrete definition remains meaningful when the area variation and quantum-entropy variation are comparable, and also where null generators enter or leave the boundary.

We make one further geometric assumption, directly analogous to the lightsheet assumption in entropy singularity theorems \cite{BoussoShahbazi2022}. The future-inward null boundary from $\sigma$ toward $B$ is assumed to be a \emph{discrete max lightsheet}. In the language of Ref.~\cite{BoussoTabor2025}, this means that the relevant portion of the edge of $C$ is future-nonexpanding in the discrete max sense, so the future lightsheet $L^+(C)$ is generated along the inward null direction. Equivalently, every genuine finite exterior cut $C_s$ reached along this lightsheet is a future-outward deformation of $C$ to which the finite-cut max-Bousso theorem applies. This condition is not derived from robust quantum trapping; the two assumptions play different roles in the proof.

\textit{Compact-Cauchy singularity theorem.--}
We now state the remaining assumptions used in the proof. The semiclassical evolution must admit valid Cauchy slices through all regions considered. If a compact Cauchy horizon forms, there must also be a valid Cauchy slice to its future. We further assume that the regulated lightsheet closes smoothly. More precisely, for nested finite exterior cuts $C_s\nearrow BC$, the regulated state and the local geometric entropy terms must approach well-defined limits. In addition, after the usual counterterms are included, removal of the regulator must be uniform near the endpoint. The End Matter proves that these assumptions ensure continuity of the renormalized generalized smooth conditional max entropy as the lightsheet closes.

We use three entropy principles in the proof. Discrete max-focusing implies the finite-cut max-Bousso bound \cite{BoussoTabor2025}. We also assume the one-shot GSL for complete causal horizons and generalized strong subadditivity, which transfers the horizon inequality to the robustly trapped wedge \cite{AkersEtAl2024}. The max-Bousso bound is applied only to genuine finite cuts of the lightsheet. It is not applied directly to the limiting wedge $BC$ because, once the lightsheet has closed, its terminal edge has disappeared and $BC$ is no longer a genuine finite cut to which the finite-cut theorem applies. Instead, the bound is established for each finite cut $C_s$ and then carried to $BC$ through the regulated limit $C_s\nearrow BC$ proved in the End Matter.

\textit{(i) One-shot GSL.} If $a\supset b$ are exterior regions associated with an earlier and a later cut of the same future causal horizon, then
\begin{equation}
 \Hgen^\eps(a|b)\leq0.
 \label{gsl}
\end{equation}
Thus every exterior cut of a complete causal horizon is past-nonexpanding.

\textit{(ii) Generalized strong subadditivity.} If $a\subset b$ and $c$ is spacelike to $b$, adjoining $c$ cannot increase the conditional generalized max entropy:
\begin{equation}
 \Hgen^\eps(b\cup_{\mathrm w}c|a\cup_{\mathrm w}c)
 \leq \Hgen^\eps(b|a).
 \label{ssa}
\end{equation}
Here $\cup_{\mathrm w}$ denotes causal, or wedge, union: the smallest domain of dependence containing both regions. The technical construction is reviewed in the End Matter.

\textit{(iii) Finite-cut max-Bousso bound.} Let $C_s$ be any genuine finite exterior cut reached by deforming $C$ along the assumed future max lightsheet $L^+(C)$. Discrete max-focusing gives
\begin{equation}
 \Hgen^\eps(C_s|C)\leq0.
 \label{finitebousso}
\end{equation}
This is the one-shot analogue of the ordinary statement that entropy on a lightsheet is bounded by its initial gravitational entropy. The endpoint $BC$ itself has no edge, so we do not insert it directly into the finite-cut theorem.

\textbf{Theorem.} \emph{Suppose that $B$ is conditionally hyperentropic relative to $C$, $\Hgen^\eps(BC|C)>0$, that the future-inward null boundary from $\sigma$ toward $B$ is a discrete max lightsheet, and that $C$ is robustly quantum trapped toward $B$. Assume discrete max-focusing, the one-shot causal-horizon GSL and generalized strong subadditivity on the relevant semiclassical evolution, together with regular regulated closure if the lightsheet closes. Then $H^+(B)$ contains a future-incomplete null geodesic.}

\textit{Proof.--} Assume, for contradiction, that every generator of $H^+(B)$ is future complete.

\textit{Step 1: no generator can remain on $H^+(B)$ forever.} Suppose that a generator $\gamma$ stays on the horizon for unbounded affine parameter. Its causal past has a null boundary $\mathcal C=\partial I^-(\gamma)$, which is a causal horizon. Let $c$ be its exterior on a Cauchy slice through the neighborhood of $\sigma$. By the one-shot GSL, $c$ is past-nonexpanding.

Join this exterior to $C$ by forming $d=C\cup_{\mathrm w}c$. The construction can be restricted to the open neighborhood used in the definition of robust trapping. It then has three properties. First, $d$ is a proper future outward null deformation of $C$ toward $B$. Second, every new part of the edge of $d$ lies on the causal horizon $\mathcal C$. Third, Eq.~\eqref{ssa} transfers past-nonexpansion from $c$ to $d$ on these new edge points. Therefore $d$ is a proper future deformation of $C$ that is past-nonexpanding everywhere on its new edge. This contradicts the assumption that $C$ is past-noncontracting. Hence a complete generator cannot remain on $H^+(B)$ for infinite affine time.

This step is where an older quantum singularity proof would require a touching lemma \cite{Wall2013}. At the classical level, if two surfaces touch and one lies inside the other, their null expansions are ordered. The analogous statement is not generally available for quantum expansions because the two shape variations probe different quantum subsystems. In our construction, the relevant edges instead coincide over an open set, so conditional strong subadditivity provides the needed comparison directly. We therefore never infer an ordering of quantum expansions merely from two surfaces touching at a single point.

\textit{Step 2: $H^+(B)$ is compact.} 
Every generator starts at the compact surface $\sigma$ and, by Step 1, leaves the horizon after a finite affine interval. These exit intervals must have a uniform upper bound. Otherwise one could choose a sequence of initial points and null directions for which the exit parameters diverge. Compactness of the initial null-normal bundle gives a convergent subsequence. The limit-curve theorem (see e.g. \cite{Hawking:1973uf,Wald:1984rg,Minguzzi:2007yq}) would then produce a generator that remains on $H^+(B)$ for arbitrarily large affine parameter, contradicting Step 1. The horizon is consequently generated from compact initial data over a bounded affine interval and is itself compact.

A generator may leave the horizon at a caustic, an intersection with another generator, or a nonsmooth junction. The proof does not need to distinguish between these possibilities. It only requires that every generator has a final point on the horizon and that the initial null data are compact. In particular, the null congruence need not remain smooth all the way to the closing set.

\textit{Step 3: finite lightsheet cuts violate the information condition at closure.} By the slicing assumption, there is a later valid Cauchy slice $\Sigma_1$ above the compact horizon, with $D^+(B)\cap\Sigma_1=\varnothing$. The lightsheet-closure lemma proved in the End Matter identifies the assumed inward max lightsheet $L^+(C)$ with the inward null component of $H^+(B)$ generated from $\sigma$, for as long as the corresponding null generators remain on the horizon. Since this component is compact and does not intersect $\Sigma_1$, we can choose a nested sequence of genuine finite cuts that approaches the closing end of the lightsheet. Their exterior wedges then satisfy $C_s\nearrow BC$. For every finite $s$, Eq.~\eqref{finitebousso} gives
\[
\Hgen^\eps(C_s|C)\leq0.
\]
The endpoint-continuity lemma proved in the End Matter shows that this finite-cut entropy approaches the entropy of the limiting wedge $BC$. Assuming the stated uniform removal of the regulator near the endpoint, one therefore obtains
\begin{equation}
 \lim_{s\to s_*^-}\Hgen^\eps(C_s|C)=\Hgen^\eps(BC|C).
 \label{endpointcontinuity}
\end{equation}
Hence $\Hgen^\eps(BC|C)\leq0$, in contradiction with conditional hyperentropy. The assumption that every generator is complete is therefore false, so at least one generator of $H^+(B)$ is future incomplete. \hfill$\square$

The proof uses compactness in a different way from Penrose's theorem. Penrose assumes a noncompact Cauchy slice and derives a topological contradiction when the null boundary generated from the trapped surface becomes compact. Here the initial Cauchy slice is compact, so that contradiction is unavailable. Instead, compactness allows the null boundary to close. Once it closes, the entropy bound contradicts the amount of independent information placed in $B$.

The conclusion is incompleteness of the semiclassical spacetime. It does not assert that curvature diverges, nor does it forbid a more fundamental quantum description beyond the endpoint. There are three possible ways for a proposed bounce to evade the conclusion: a null generator may end within the effective geometry; the valid semiclassical slicing may fail before reaching the bounce; or one of the assumed entropy laws may fail. In each case, a regular bounce cannot be established entirely within the same controlled semiclassical evolution while keeping all null generators complete.

This distinction is important for effective models of singularity resolution. A metric can be smooth in a chosen coordinate chart while the quantum state, the semiclassical expansion, or the effective gravitational action has already left its regime of validity. Conversely, the theorem does not identify such a breakdown as a curvature singularity. It identifies the point at which a complete bounce can no longer be inferred from semiclassical geometry and the stated entropy laws.

\textit{Closed Friedmann universe.--}
Consider a closed Friedmann--Lema\^itre--Robertson--Walker (FLRW) geometry,
\begin{equation}
 ds^2=-dt^2+a^2(t)\left(d\chi^2+\sin^2\!\chi\,d\Omega_2^2\right).
 \label{flrw}
\end{equation}
The areal radius of a symmetry sphere is $R=a\sin\chi$. Future radial null vectors may be normalized as $k_\pm=\partial_t\pm a^{-1}\partial_\chi$, where the two signs point toward increasing and decreasing $\chi$. Since the area is $4\pi R^2$, its fractional rate of change along these vectors gives
\begin{equation}
 \theta_\pm=\frac{k_\pm(4\pi R^2)}{4\pi R^2}
 =2\left(H\pm\frac{\cot\chi}{a}\right),\qquad H=\frac{\dot a}{a}.
 \label{expansions}
\end{equation}
At the equator, $\chi=\pi/2$, both expansions equal $2H$. Hence the equatorial sphere is classically trapped throughout any contracting epoch, $H<0$: future-directed light rays initially shrink in area whichever side they enter. A strictly contracting classical surface is stable under small deformations. Thus, when quantum corrections to the generalized entropy are controlled and do not reverse the sign, either hemisphere provides the robustly trapped region $C$ required by the theorem.

The area of the equator is $A_\sigma=4\pi a^2$. The exact information condition for the opposite hemisphere $B$ is therefore
\begin{equation}
 \Hmax^\eps(B|C)>\frac{\pi a^2}{G\hslash}+O(G\hslash),
 \label{flrwbound}
\end{equation}
where the correction includes the local terms already described below Eq.~\eqref{generalizedmax}. This is a statement about the quantum state on the full slice, not about the FLRW metric alone. In particular, the energy density and scale factor do not determine the correlations between $B$ and $C$, and therefore do not determine the conditional entropy $\Hmax^\eps(B|C)$.

For illustration, consider a homogeneous mixed state whose correlations across the equator are short compared with $a$. A hemisphere has volume
\begin{equation}
 V_B=4\pi a^3\int_0^{\pi/2}\!d\chi\,\sin^2\chi=\pi^2a^3.
\end{equation}
If $s_{\max}$ is the coarse-grained max-entropy density, then, up to subleading boundary correlations,
\[
\Hmax(B|C)\simeq V_B s_{\max}.
\]
Equation~\eqref{flrwbound} therefore gives
\begin{equation}
s_{\max}>\frac{1}{\pi G\hslash a},
\label{densitybound}
\end{equation}
up to smoothing, correlation, and higher-curvature corrections. The key point is the scaling: the bulk information grows as $a^3$, while the gravitational bound grows only as $a^2$.
Nevertheless, Eq.~\eqref{densitybound} is only an intuitive estimate. The theorem uses the exact conditional quantity in Eq.~\eqref{flrwbound}, which remains applicable when the state is strongly correlated and no local entropy density exists.

The information condition is essential. A trapped surface by itself does not rule out a bounce on a compact slice, as global de Sitter space demonstrates. For a globally pure state, smooth-entropy duality gives
\begin{equation}
 \Hmax^\eps(B|C)=-\Hmin^\eps(B),
\end{equation}
so the conditional max entropy is not generically positive
\cite{TomamichelColbeckRenner2010,Tomamichel2016}. The de Sitter vacuum therefore need not satisfy Eq.~\eqref{flrwbound}. This is not a loophole: the theorem excludes closed cosmologies only when contraction is accompanied by sufficiently large independent information in $B$.

The FLRW example also shows how the theorem can be tested in practice. One chooses a time in a controlled contracting regime and checks three things: robust trapping of the equator, existence of the future-inward discrete max lightsheet toward the chosen hemisphere, and the conditional hyperentropy bound. In the smooth regime with strictly negative future expansion, the lightsheet condition follows when finite-$G\hslash$ corrections do not reverse the discrete max-nonexpansion sign. Only when these geometric and information conditions hold does incompleteness follow. It can be noted that homogeneity simplifies the test but is not assumed by the theorem.

\textit{Discussion.--}
The result bridges the gap between entropy-based singularity theorems on compact slices and the robust finite-$G\hslash$ singularity theorem. Its global structure closely follows the hyperentropic-region argument: excess information is combined with the existence of an inward lightsheet. At finite $G\hslash$, two additional ingredients are needed. The lightsheet is defined through discrete max-nonexpansion, while robust quantum trapping excludes a complete generator that could remain indefinitely on the Cauchy horizon. Conditional hyperentropy then provides the obstruction when the lightsheet closes. The theorem therefore does not derive the lightsheet condition from robust trapping; the two assumptions play distinct roles. A nonsingular closed bounce can evade the theorem if the conditional hyperentropy condition fails, if the required inward max lightsheet does not exist, if the evolution leaves the semiclassical regime, or if one of the discrete entropy laws used in the proof is violated.

The main dynamical question is now concrete: can controlled matter states satisfy Eq.~\eqref{flrwbound} while all curvature scales remain below the effective-theory cutoff? Entropy bounds have already been tested in explicit homogeneous nonsingular geometries \cite{KanaiNomuraYoshida2023}. Establishing the conditional max-entropy inequality in a microscopic closed-universe model would turn the theorem into a direct test of that model.

The result is state dependent. Rather than replacing the geometric assumptions with a single universal entropy criterion, it asks whether a given state contains more independent information than the contracting boundary can support. This distinction is important in quantum gravity, since different quantum states on the same smooth geometry can have very different correlation structures. The theorem also provides a concrete test for nonsingular bounce models: one may either construct a valid semiclassical continuation in which every trapped partition remains below the conditional entropy bound, or identify which of the discrete focusing assumptions fails.

\bibliography{refs}
\newpage
\onecolumngrid
\section*{End Matter}
\twocolumngrid

This End Matter contains the formal causal constructions and entropy-composition statements used in the Letter. The physical motivation, theorem, full proof, and cosmological application are contained in the main text.

\section{Wedges and localized null deformations}

For a set $X$ in a globally hyperbolic spacetime, let $I(X)=I^+(X)\cup I^-(X)$ and define its spacelike complement by
\begin{equation}
 X'=\Int[M\setminus I(X)].
 \label{sm:complement}
\end{equation}
A wedge is an open region $a$ satisfying $a=a''$. A spatial region and its domain of dependence define the same wedge, so the construction does not depend on which Cauchy slice represents the region~\cite{BoussoTabor2025}. The edge is the codimension-two set
\begin{equation}
 \partial a\setminus I(a).
\end{equation}
For two wedges, their wedge union is the smallest wedge containing both:
\begin{equation}
 a\mathbin{\mathop{\cup}\limits_{\rm w}}b=(a'\cap b')'.
 \label{sm:wedgeunion}
\end{equation}
This operation, rather than an ordinary set union on one time slice, is required when the two regions are represented on different slices~\cite{BoussoTabor2025}.

The proof in the Letter uses the following localized construction. Let $C$ be the exterior of the initial surface $\sigma$, and suppose a future-complete generator $\gamma$ remains on $H^+(B)$. Its past causal horizon
\begin{equation}
 \mathcal C=\partial I^-(\gamma)
 \label{sm:causalhorizon}
\end{equation}
has an exterior wedge $c$. Choose a small open neighborhood $O$ of the portion of $\sigma$ approached by $\mathcal C$ and restrict $c$ so that it agrees with $C$ outside $O$. Then
\begin{equation}
 d=C\mathbin{\mathop{\cup}\limits_{\rm w}}c
 \label{sm:deformation}
\end{equation}
is a proper future outward null deformation of $C$ supported in $O$. Every new edge point of $d$ is inherited from the relevant cut of $\mathcal C$. One does not need a claim about points outside $O$.

This localization matters at finite $G\hslash$. A classical maximum principle compares the expansions of two smooth null surfaces that touch. A quantum expansion, however, contains an entropy variation and therefore depends on the quantum subsystem selected by the entire cut. Distinct cuts that touch at only one point need not define comparable subsystems. Equation~\eqref{sm:deformation} instead produces nested wedges that share an open edge portion. Their generalized conditional entropies can be compared by strong subadditivity~\cite{BoussoTabor2025}.

\section{Entropy inequalities used by the construction}

For nested wedges $a\subset b$, the generalized smooth conditional max entropy is the renormalized combination
\begin{equation}
 \Hgen^\eps(b|a)=
 \frac{A(b)-A(a)}{4G\hslash}
 +\Hmax^\eps(b|a)+S_{\rm ct}(b|a).
 \label{sm:hgen}
\end{equation}
Here $S_{\rm ct}$ contains the local terms fixed by the gravitational effective action. The separate area, matter, and counterterm contributions depend on the regulator, while their sum does not. Smoothing allows the state to vary within a distance $\eps$ of the original state, which makes the entropy stable under small perturbations~\cite{Tomamichel2016}. Importantly, $\Hmax^\eps$ is a one-shot quantity: it is defined for the single state under consideration and does not require introducing many independent copies of the state or taking a thermodynamic or many-copy limit.

The transfer of past-nonexpansion from $c$ to $d$ uses generalized strong subadditivity~\cite{BoussoTabor2025}. If $a\subset b$ and $C$ is spacelike to $b$, then
\begin{equation}
 \Hgen^\eps\!\left(
 b\mathbin{\mathop{\cup}\limits_{\rm w}}C
 \middle|
 a\mathbin{\mathop{\cup}\limits_{\rm w}}C
 \right)
 \leq \Hgen^\eps(b|a).
 \label{sm:ssa}
\end{equation}
Apply this inequality to each sufficiently small past outward enlargement\footnote{Here an ``enlargement'' means a small outward null deformation of the wedge: a portion of its edge is moved slightly outward along the chosen past-directed null direction, producing a nearby larger wedge.} $b\supset c$ of the horizon exterior $a=c$. The right-hand side is nonpositive by the assumed one-shot generalized second law~\cite{AkersEtAl2024}. After adjoining $C$ by wedge union, the same deformation gives the corresponding enlargement of $d=C\cup_{\rm w}c$. Equation~\eqref{sm:ssa} therefore shows that this enlargement also has nonpositive generalized conditional max entropy. Hence $d$ is past-nonexpanding at every new edge point. This is precisely the type of deformation ruled out by robust quantum trapping~\cite{Bousso2025}.

The global step is formulated only on genuine finite cuts of the assumed max lightsheet. Let $C_s$ denote the exterior wedge at a finite cut, with
\begin{equation}
 C\subset C_s\subset BC,\qquad \partial C_s\subset \partial C\cup L^+(C).
 \label{sm:finitecuts}
\end{equation}
The finite-cut max-Bousso theorem, which is a direct consequence of discrete max-focusing~\cite{BoussoTabor2025}, gives
\begin{equation}
 \Hgen^\eps(C_s|C)\leq0
 \label{sm:finitebousso}
\end{equation}
for every such finite cut. This inequality compares each $C_s$ directly with the initial wedge $C$, so the argument does not require a sequence of intermediate entropy inequalities to be combined along the lightsheet. Moreover, the limiting wedge $BC$ is not a genuine finite cut because the boundary surface associated with the cut has disappeared. The finite-cut theorem is therefore applied only to $C_s$, and the result for $BC$ is obtained afterward by taking the regulated limit $C_s\nearrow BC$.

\textbf{Lemma (lightsheet closure from the compact Cauchy horizon).} \emph{Let $\Sigma=B\cup\sigma\cup C$ be a Cauchy slice and let $L^+(C)$ be the future-outward null boundary of $C$ directed through $\sigma$ toward $B$. Before its generators leave the achronal boundary, $L^+(C)$ is the same null hypersurface as the inward component of $H^+(B)$ generated from $\sigma$. If this component of $H^+(B)$ is compact and there exists a Cauchy slice $\Sigma_1$ to its future with $D^+(B)\cap\Sigma_1=\varnothing$, then $L^+(C)$ closes and possesses a cofinal\footnote{Here ``cofinal'' simply means that the cuts can be taken arbitrarily close to the endpoint of the closing lightsheet.} nested family of genuine finite cuts whose exterior wedges obey $C_s\nearrow BC$.}

\emph{Proof.} On the initial slice, $B$ and $C$ share the edge $\sigma$. The future null generators orthogonal to $\sigma$ and directed from $C$ into $B$ form the local future boundary separating points whose past meets $B$ from those remaining in the exterior of $B$. Hence this null boundary is simultaneously the relevant future-outward boundary of $C$ and the inward component of $\partial D^+(B)=H^+(B)$, until a generator reaches its last point on the achronal boundary. Generator exits at caustics, crossovers, or nonsmooth junctions do not alter the equality of the two boundaries on the portion where they are generated from $\sigma$~\cite{Hawking:1973uf,Wald:1984rg}.

If that component of $H^+(B)$ is compact, all of its generators have last horizon points in a compact set. A Cauchy slice $\Sigma_1$ lying strictly to its future and satisfying $D^+(B)\cap\Sigma_1=\varnothing$ lies beyond every such last point. Choose any increasing sequence of cuts that approaches the last-point set from below along the null generators, avoiding the exit set at each finite stage. The associated exterior wedges are genuine finite deformations of $C$, are nested, and exhaust every point not belonging to the vanishing future development of $B$. Since no point of $D^+(B)$ remains on $\Sigma_1$, their wedge limit is the full closed-slice wedge $BC$. Therefore $C_s\nearrow BC$. \hfill$\square$

\textbf{Lemma (regulated endpoint continuity).} 
\emph{Let $\{C_s\}_{s<s_*}$ be a nested family of genuine finite exterior cuts of a closing max lightsheet, with $C_s\nearrow BC$ as $s\to s_*^-$. Let $\Lambda$ denote a UV regulator and fix a smoothing parameter $\eps$. Assume that: (i) for each fixed $\Lambda$, the regulated conditional state data for $C_s|C$ and $BC|C$ are represented in a common finite-dimensional algebra, and the former approach the latter in purified distance as $s\to s_*^-$ (that is, the regulated quantum states become arbitrarily close in the standard smooth-entropy metric); (ii) the terminal edges $\sigma_s=\partial C_s$ close regularly, with $A(\sigma_s)\to0$; (iii) all local geometric counterterm densities entering the generalized entropy remain bounded along the closing family; and (iv) after the standard local counterterms are included, removal of the regulator is uniform for $s$ in a one-sided neighborhood of $s_*$, including the endpoint. Then the renormalized generalized conditional entropy is continuous at closure,}
\begin{equation}
 \lim_{s\to s_*^-}\Hgen^\eps(C_s|C)
 =\Hgen^\eps(BC|C).
 \label{sm:endpointlimit}
\end{equation}

\emph{Proof.} At fixed regulator $\Lambda$, write the regulated generalized conditional entropy as the sum of its regulated matter term and local geometric terms,
\begin{align}
 H_{\max,\mathrm{gen},\Lambda}^{\eps}(C_s|C)
 &=\frac{A(\sigma_s)-A(\sigma)}{4G\hslash}
 \nonumber\\[-2pt]
 &\quad+H_{\max,\Lambda}^\eps(C_s|C)+S_{{\rm ct},\Lambda}(C_s|C).
 \label{sm:regulateddecomp}
\end{align}
The conditioning edge $\sigma=\partial C$ is fixed. By regular closure $A(\sigma_s)\to0=A[\partial(BC)]$, so the area contribution tends to its endpoint value. Likewise each local counterterm on the terminal edge has the form $\int_{\sigma_s}\sqrt h\,{\cal I}_{\Lambda}$, with $|{\cal I}_{\Lambda}|$ bounded along the closing family at fixed $\Lambda$. Hence its absolute value is bounded by a constant times $A(\sigma_s)$ and vanishes as $s\to s_*^-$. The fixed-edge counterterms are unchanged.

It remains to control the matter term. Let $\rho_{s,\Lambda}$ and $\rho_{*,\Lambda}$ denote the regulated conditional state data for $C_s|C$ and $BC|C$ in the common finite-dimensional algebra. By assumption $P(\rho_{s,\Lambda},\rho_{*,\Lambda})\to0$, where $P$ is purified distance~\cite{Tomamichel2016}. For fixed finite dimension the unsmoothed conditional max entropy is continuous in the state, and the closed $\eps$-ball $\mathcal B_\eps(\rho)=\{\widetilde\rho:P(\widetilde\rho,\rho)\le\eps\}$ is compact~\cite{Tomamichel2016}. Thus the optimized value $H_{\max,\Lambda}^\eps$ is continuous in the center of the smoothing ball. Explicitly, choose approximate optimizers $\widetilde\rho_s\in\mathcal B_\eps(\rho_s)$. Compactness gives a convergent subsequence $\widetilde\rho_{s_n}\to\widetilde\rho_*$. Continuity of $P$ implies $\widetilde\rho_*\in\mathcal B_\eps(\rho_*)$, while continuity of the unsmoothed entropy gives the corresponding lower-semicontinuity bound. Conversely, choose an endpoint approximate optimizer in the interior of an arbitrarily slightly enlarged smoothing ball,
\[
P(\hat{\rho}_*,\rho_*)<\varepsilon+\delta .
\]
For sufficiently late $s$, the triangle inequality gives
\[
P(\hat{\rho}_*,\rho_s)<\varepsilon+2\delta .
\]
Letting first $s\to s_*^-$ and then $\delta\to0$ yields the opposite bound. In finite dimension, the optimized smooth conditional max entropy is continuous in the smoothing radius at fixed $\varepsilon$~\cite{Tomamichel2016}. Hence the limit $\delta\to0$ returns the optimization over the original $\varepsilon$-ball and gives the required opposite semicontinuity bound. Therefore
\begin{equation}
\lim_{s\to s_*^-}
H^\varepsilon_{\max,\Lambda}(C_s|C)
=
H^\varepsilon_{\max,\Lambda}(BC|C).
\end{equation}
Combining the matter limit with the area and counterterm limits gives, for every fixed $\Lambda$,
\begin{equation}
 \lim_{s\to s_*^-}H_{\max,\mathrm{gen},\Lambda}^{\eps}(C_s|C)=H_{\max,\mathrm{gen},\Lambda}^{\eps}(BC|C).
 \label{sm:fixedreglimit}
\end{equation}
Assumption (iv) now permits the regulator limit to be interchanged with the one-sided closing limit. Writing $\Hgen^\eps=\lim_{\Lambda\to\infty}H_{\max,\mathrm{gen},\Lambda}^\eps$ after the local counterterms are included, uniform convergence near $s_*$ gives
\begin{equation}
 \lim_{s\to s_*^-}\Hgen^\eps(C_s|C)=\Hgen^\eps(BC|C),
\end{equation}
which proves Eq.~\eqref{sm:endpointlimit}. \hfill$\square$

The point of assumption (iv) is only to control the order of limits; no regulator-independent continuity theorem for arbitrary gravitational wedge algebras is assumed. The finite-cut max-Bousso inequality is an inequality for the renormalized generalized entropy. Once the renormalized continuity above is established, taking $s\to s_*^-$ in Eq.~\eqref{sm:finitebousso} gives $\Hgen^\eps(BC|C)\leq0$. Thus the endpoint step uses the finite-cut theorem and regulator removal in a definite order, with no direct application of the Bousso bound to the empty-edge wedge.

\section{Compactness and the empty future slice}

For completeness, we spell out the causal compactness step used in the Letter. The future horizon $H^+(B)$ is ruled by null geodesics starting orthogonally from the compact edge $\sigma$, up to the usual nonsmooth generator junctions. Suppose every generator is complete but no generator remains on the horizon forever. Parameterize the initial null data by a compact set $K$ and let $\lambda_*(q)<\infty$ be the last affine parameter for which the generator from $q\in K$ lies on $H^+(B)$.

If the exit parameters were not uniformly bounded, there would be a sequence $q_n\to q\in K$ with $\lambda_*(q_n)\to\infty$. For each fixed $L$, the corresponding generator segments of length $L$ have a limit segment contained in the closed achronal boundary $H^+(B)$. A diagonal limit as $L\to\infty$ gives a future-inextendible generator from $q$ that remains on the horizon~\cite{Minguzzi:2007yq,Hawking:1973uf}. Under the assumed completeness this generator has unbounded affine length, contradicting the result of the localized deformation argument. Hence $\lambda_*$ is uniformly bounded.

The horizon is contained in the image of compact initial data evolved over this bounded parameter interval and is closed, so it is compact. Global hyperbolicity then permits a Cauchy slice $\Sigma_1$ lying strictly to its future~\cite{Hawking:1973uf,Wald:1984rg}. We claim that
\[
D^+(B)\cap\Sigma_1=\varnothing .
\]
Suppose instead that
\[
A_1:=D^+(B)\cap\Sigma_1
\]
is nonempty. Since $B$ is a proper region of the initial Cauchy slice $\Sigma$, choose a point $c$ in the interior of $C$. A timelike curve through $c$ meets $\Sigma_1$ at some point $p_c$, and its past-inextendible extension meets the initial slice at $c\notin B$. Hence $p_c\notin D^+(B)$ by the definition of the future domain of dependence. Therefore $A_1$ is a proper subset of $\Sigma_1$.

For a closed initial region in a globally hyperbolic spacetime, $D^+(B)$ is closed~\cite{Hawking:1973uf,Wald:1984rg}. Thus $A_1$ is closed in $\Sigma_1$. Since $\Sigma_1$ is connected and $A_1$ is assumed nonempty and proper, its boundary in $\Sigma_1$ is nonempty. Let
\[
q\in\partial_{\Sigma_1}A_1 .
\]
Then $q$ is a boundary point of $D^+(B)$ lying strictly to the future of the initial slice. By the standard boundary structure of a domain of dependence, such a point lies on the future Cauchy horizon,
\[
q\in H^+(B).
\]
But $\Sigma_1$ was chosen strictly to the future of the compact set $H^+(B)$, and therefore
\[
H^+(B)\cap\Sigma_1=\varnothing ,
\]
which is a contradiction. Hence
\begin{equation}
D^+(B)\cap\Sigma_1=\varnothing .
\end{equation}
The lightsheet-closure lemma then turns this empty future development into a cofinal family of finite cuts of the same inward null boundary used in Eq.~\eqref{sm:finitebousso}, with $C_s\nearrow BC$. The additional assumptions in the Letter are that this slice remains within the valid semiclassical regime and that the regulated closing family is regular, including the uniform regulator-removal condition required for Eq.~\eqref{sm:endpointlimit}; global hyperbolicity alone cannot guarantee either property of an effective theory.

\section{One-shot duality for a pure closed universe}

The de Sitter comment in the Letter uses a standard smooth-entropy identity~\cite{TomamichelColbeckRenner2010,Tomamichel2016}. If $BCR$ purifies a state on $BC$, smooth entropy duality gives, with the corresponding smoothing convention,
\begin{equation}
 \Hmax^\eps(B|C)=-\Hmin^\eps(B|R).
 \label{sm:duality}
\end{equation}
For a pure state of the complete closed universe, the purifying system $R$ is trivial. Thus the conditional max entropy of a hemisphere is not its ordinary thermodynamic entropy and need not be positive. Geometric trapping in the contracting half of global de Sitter therefore does not imply the conditional hyperentropy assumption. This illustrates why the theorem is state dependent even in a fixed background geometry.

\end{document}